\documentclass[aps,pdflatex,superscriptaddress,floats,showpacs,10pt]{revtex4-1}
\usepackage{graphicx}
\usepackage{bm}
\usepackage{amssymb}
\usepackage{amsmath}

\begin{document}

\title{Attractors and chaos of electron dynamics in electromagnetic standing wave}
\author{Timur Zh. Esirkepov}
\affiliation{QuBS, Japan Atomic Energy Agency, Kizugawa, Kyoto 619-0215, Japan}
\author{Stepan S. Bulanov}
\affiliation{University of California, Berkeley, CA 94720, USA}
\author{James K. Koga}
\author{Masaki Kando}
\author{Kiminori Kondo}
\affiliation{QuBS, Japan Atomic Energy Agency, Kizugawa, Kyoto 619-0215, Japan}
\author{Nikolay N. Rosanov}
\altaffiliation[Also at ]{the ITMO University, Saint-Petersburg 197101, Russia}
\affiliation{Vavilov State Optical Institute, Saint-Petersburg 199034, Russia}
\author{Georg Korn}
\affiliation{ELI Beamline Facility, Institute of Physics, Czech Academy of Sciences, Prague 18221, Czech Republic }
\author{Sergei V. Bulanov}
\altaffiliation[Also at ]{A. M. Prokhorov Institute of General Physics of RAS, Moscow, Russia}
\altaffiliation{the ITMO University, Saint-Petersburg 197101, Russia}
\affiliation{QuBS, Japan Atomic Energy Agency, Kizugawa, Kyoto 619-0215, Japan}


\begin{abstract}
The radiation reaction radically influences
the electron motion in an electromagnetic standing wave
formed by two super-intense counter-propagating laser pulses.
Depending on the laser intensity and wavelength,
either classical or quantum mode of radiation reaction prevail, or both are strong.
When radiation reaction dominates,
electron motion evolves to limit cycles and strange attractors.
This creates a new framework for high energy physics experiments
on an interaction of
energetic charged particle beams and colliding super-intense laser pulses.

\end{abstract}


\maketitle


New regimes of light-matter interaction
emerge with the increase of laser power beyond petawatt,
leading to radiation pressure dominant acceleration of ions,
bright coherent x-ray generation and electron-positron pairs creation \cite{MTB}.
The next generation of high power short-pulse lasers  will soon reach
the intensity of electromagnetic (EM) radiation
of $10^{23}$ W/cm$^{2}$ for a $1\,\mu$m wavelength
\cite{ELI-Beamlines},
$10^5$ times greater than the relativistically strong intensity
threshold $I_0 =1.37\times 10^{18}$ W/cm$^{2}$.
Then the EM radiation emission by electrons
will be substantial \cite{MTB, MaShu, DPHK},
making the electron dynamics strongly dissipative \cite{ZKS, KEB}
and causing the laser energy fast conversion to hard EM radiation,
in the gamma-ray spectral range for typical laser parameters \cite{Ridgers, NakaKo}.

The laser intensity above $10^{23}$ W/cm$^{2}$ brings novel physics 
\cite{BELL} 
(see \cite{MTB, MaShu, DPHK, Fedotov, SSB, SSB-2013, AGZ-2014, ELKINA-2014, VRANIC-2014}
for details), 
where the electron (positron) dynamics is principally determined by
the radiation reaction (RR) in its classical form or
quantum electrodynamics (QED) effects. 
The latter weaken the EM emission of relativistic electrons,
thus decreasing the classical RR \cite{JSCHW, AAS}.

Even in a simple case of a standing wave (SW),
the electron dynamics is surprisingly complicated.
At the magnetic field node plane,
the electron motion is unstable \cite{SSB}, with the instability growth rate 
being approximately equal to the EM field frequency.
In a linearly polarized SW, the electrons are concentrated
in the SW spatial periods \cite{JIPUKHOV, Gonoskov, NEITZ}.
The SW configuration is widely used 
in the QED theory of superstrong EM field interaction with vacuum and charged particles, 
because the processes description is greatly simplified at 
the magnetic field node plane:
there the electric field vector merely rotates 
in a circularly polarized (CP) or oscillates
in a linearly polarized (LP) SW,
with nonzero Poincar\'{e} invariants.
In addition, the SW formed by 
two perfectly matched counter-propagating laser pulses 
has a two times higher electric field
than a single laser pulse,
which facilitates QED effects as in the multi-beam configuration \cite{SSB-multibeam}.

In this Letter we show that
the SW intensity and wavelength are uncoupled critical parameters determining
the electron dynamics in a strong EM SW.
The intensity--wavelength plane is divided into four domains where 
the RR is either negligible,
or substantially classical,
or mainly manifests itself as a QED effect while the classical RR force is small,
or the classical RR force and QED effects are both strong.
When the RR is significant,
a strongly dissipative motion of electrons
results in formation of limit cycles and strange attractors.


The electron dynamics in the EM wave is described by the equation
\begin{equation}
\dot{\bf p}=e ({\bf E}+\bm{\beta} \times {\bf B})+{\bf f}_{\rm rad}
,\,\,
{\bf f}_{\rm rad} = G_e {\bf f}_{\rm LL} ,
\label{EquMot}
\end{equation}
where 
${\bf p}=m_e c \gamma_e\boldsymbol{\beta}$,
$\boldsymbol{\beta}={\bf v}/c$,
$\gamma_e=(1-\beta^2)^{-1/2}$;
${\bf E}$ and ${\bf B}$ are the electric and magnetic fields;
$e$, $m_e$, ${\bf v}$ and ${\bf p}$ are 
the electron charge, mass, velocity and momentum;
$c$ is the speed of light in vacuum.
The RR force in the Landau--Lifshitz form, ${\bf f}_{\rm LL}$ \cite{LL-TP},
is reduced by a factor $G_e$ representing
the classical RR weakening due to QED effects,
following the approach of Refs. \cite{BELL, Ridgers, SSB-2013}.

In the ultrarelativistic limit $\gamma_e \gg 1$,
the RR force can be written as 
\begin{equation}
{\bf f}_{\rm rad} \approx -\varepsilon_{\rm rad} G_e m_e c \omega  \boldsymbol{\beta} a_S^2 \chi_e^2.
\label{eq:frad}
\end{equation}
Here 
$\varepsilon_{\rm rad} = 
4 \pi r_e/3 \lambda
\approx 1.18\times 10^{-8} (1 \mu{\rm m}/\lambda)$;
$r_e=e^2/m_e c^2 \approx 2.82\times 10^{-13}$cm is the classical electron radius;
$a_S=e E_S/m_e\omega c
\approx 4.12\times 10^{5}(\lambda/1 \mu {\rm m})$
corresponds to the QED critical field 
$E_S=
\alpha e/r_e^2$;
$\alpha = e^2/\hbar c$ is the fine-structure constant
\cite{BLP};
$\omega$ and $\lambda$ are the EM wave frequency and wavelength.
The relativistic and gauge invariant parameter
\begin{equation}
\chi_e=(\gamma_e/E_S)
[ ({\bf E}+\boldsymbol{\beta}\times {\bf B})^2 - (\boldsymbol{\beta}\cdot {\bf E})^2 ]^{1/2}
\label{eq:chie}
\end{equation}
characterizes the probability of a gamma-photon emission
by an electron with momentum ${\bf p}$.
QED effects are negligible for $\chi_e\ll 1$ and 
become substantial for $\chi_e\simeq 1$.

The more significant QED effects in the electron motion, the less radiation is emitted.
According to \cite{JSCHW} the total radiated power is reduced by 
a factor depending on the parameter $\chi_e$. 
Introduced in Eq. (\ref{eq:frad}) this factor is written
using \cite{BLP, RITUS,  IVS} as
\begin{equation}
G_e(\chi_e)=-
\int^{\infty}_0
\frac{3+1.25 \chi_e \xi^{3/2}+3 \chi_e^2 \xi^3}{\left(1+\chi_e \xi^{3/2}\right)^4}
{\rm A\!i}^{\prime}(\xi) \xi d\xi,
\label{G-chi-e}
\end{equation}
where ${\rm A\!i}(x)$ is the Airy function. The photon emission discreet nature
is neglected (see \cite{DUCLOUS, BRADY, SSB-2013}).
For computations we approximate Eq.(\ref{G-chi-e}) by
$G_e(\chi_e)\approx(1 +18\chi_e +69\chi_e^2 + 73\chi_e^3 + 5.806\chi_e^4)^{-1/3}$,
accurate within $10^{-3}$ for $0<\chi_e<10$,
with the same asymptotic at 0 and $\infty$ as Eq. (\ref{G-chi-e}).


We consider the electric field of the one-dimensional (1D) 
CP EM SW near the magnetic field node plane:
$\bm{a}=-a ({\mathfrak i}_2 \cos \tau + {\mathfrak i}_3 \sin \tau )$,
where ${\mathfrak i}_2$ and ${\mathfrak i}_3$ are orthogonal unit vectors
perpendicular to the SW axis; 
$\tau=\omega t, \quad {\bf q}={\bf p}/{m_e c}$, and
$a=e E/m_e \omega c = (I/I_0)^{1/2}(\lambda/1 \mu {\rm m})$.
We change to the rotating coordinate system
\cite{LADvsLL},
\begin{equation}
q_\parallel = q_2 \cos\tau + q_3 \sin\tau , \,\,
q_\perp = q_2 \sin\tau - q_3 \cos\tau .
\label{qRot}
\end{equation}
Eq. (\ref{eq:chie}) yields
$\chi_e=(a/a_S)(1+q_{1}^2+q_{\perp}^2)^{1/2}$.

Substituting Eq. (\ref{qRot}) into Eq. (\ref{EquMot}) and
neglecting  the electron momentum along the SW axis,
$q_1\ll (q_2^2+q_3^2)^{1/2}$,
we obtain
\begin{eqnarray}
\dot q_{||}+q_{\perp}=a
	 -\varepsilon_{\rm rad}\, G_e(\chi_e) a^2 q_{||}q_{\perp}^2/\gamma_e,
\label{EquMotQpar}
\\
\dot q_{\perp}-q_{||}=
	 -\varepsilon_{\rm rad}\,G_e(\chi_e)[\gamma_e a+a^2 q_{\perp}(1+ q_{\perp}^2)/\gamma_e],
\label{EquMotQprp}
\end{eqnarray}
where the dot denotes differentiation with respect to $\tau$.
Solutions of this system 
asymptotically tend to steady state (provided $\varepsilon_{\rm rad}>0$).
Following \cite{KEB}, from Eqs. (\ref{EquMotQpar}-\ref{EquMotQprp})
we find the critical EM amplitude
determining the RR strength,
\begin{equation}
a_{\rm RQ} = [\varepsilon_{\rm rad}G_e(\chi_e)]^{-1/3}.
\label{aRQ}
\end{equation}
RR is negligible for $a\ll a_{\rm RQ}$ and 
becomes substantial for $a\simeq a_{\rm RQ}$.
In this limit,
for $q_\perp \propto a_{\rm RQ}$
with a factor of the order of unity,
we estimate the $\chi_e$ parameter
as $\chi_m \approx a_{\rm RQ} q_\perp/a_S \approx \eta a_{\rm RQ}^2/a_S$
corresponding to $G_m = G_e(\chi_m)$.
This gives the critical wavelength at which RR becomes substantially quantum,
\begin{equation}
\lambda_{\rm RQ} = 9\pi r_e/2 \alpha^3 \chi_m^3 G_m^2.
\label{lambdaRQ}
\end{equation}
For $\chi_m=1$ and $G_m\approx 0.18$,
we obtain $I_{\rm RQ} = a_{\rm RQ}^2 I_0 = 1.75\times 10^{24}$ W/cm$^2$
and $\lambda_{\rm RQ} = 3.1\,\mu$m.

\begin{figure}[H]
\includegraphics{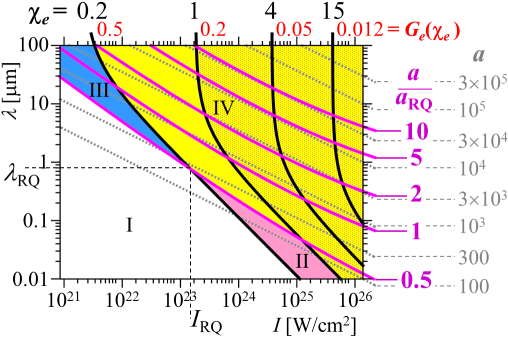}
\caption{%
(color).
Stationary motion of electron in a rotating electric field with the
dimensionless amplitude $a$ (gray dashed curves):
$\chi_e$ parameter (black curves) and 
corresponding factor $G_e(\chi_e)$ (red), and 
the amplitude normalized to 
$a_{\rm RQ}=[\varepsilon_{\rm rad}G_e(\chi_e)]^{-1/3}$ (magenta curves),
versus the electric field intensity, $I$, and wavelength, $\lambda$.
Radiation reaction (RR) is negligible in the region I,
QED effects dominate while the classical RR force is small in II,
RR is mostly classical in III,
the classical RR force and QED effects are both strong in IV.
Hatched region where discrete nature of electron emission cannot be neglected.
}
\label{fig-1:I-lambda-zones}
\end{figure}

Taking $d/d\tau=0$ in Eqs. (\ref{EquMotQpar}-\ref{EquMotQprp}), 
corresponding to the steady state solution \cite{LADvsLL},
we obtain algebraic dependences of 
$q_\parallel, \gamma_e$ on $q_\perp$,
and the expression
\begin{equation}
a-q_\perp = [\varepsilon_{\rm rad} G_e(\chi_e)]^2 a^2 q_\perp^3 (1+a q_\perp),
\label{qper}
\end{equation}
implicitly defining $q_\perp$ as a function of $a$, $\varepsilon_{\rm rad}$, and $a_S$.
Thus all the dependent variables can be represented as functions of the EM SW intensity,
$I=E^2c/4\pi$, and wavelength, $\lambda$.
In this way Fig. \ref{fig-1:I-lambda-zones}
shows the $\chi_e$ parameter,
corresponding factor $G_e(\chi_e)$,
and the EM SW amplitude normalized to $a_{\rm RQ}$.

Fig. \ref{fig-1:I-lambda-zones}
reveals domains of different role of RR in
the model of the electron stationary motion in a rotating electric field.
The classical RR effects becomes substantial when $a\gtrapprox 0.5 a_{\rm RQ}$
while QED effects come into play at $\chi_e \gtrapprox 0.2$,
which corresponds to the classical RR weakening with the factor $G_e\lessapprox 0.5$.
The intersection of the curves $a/a_{\rm RQ}=0.5$ and $\chi_e=0.2$
gives the characteristic intensity
$I_{\rm RQ}^* \approx 1.5\times 10^{23}$W/cm$^2$,
and wavelength
$\lambda_{\rm RQ}^* \approx 0.76\,\mu$m,
within the order of magnitude of the estimate presented above.
This is a meeting point of four different domains:
RR is negligible in the region I,
QED effects dominate while the classical RR force is small in II,
RR is mostly classical in III,
the classical RR force and QED effects are both strong in IV.
Beyond $\chi_e\gtrapprox 1$, a discrete nature of electron emission cannot be neglected
posing the applicability limit for the model.


In order to generalize the picture given by Eqs. (\ref{EquMotQpar}-\ref{EquMotQprp})
and investigate the electron dynamics with RR in classical and QED modes,
we numerically simulate the electron motion in a 1D EM SW according to 
Eqs. (\ref{EquMot}), (\ref{eq:chie}), (\ref{G-chi-e}).
In our setting the electric field field oscillates at antinodes,
$x = \pm{\mathfrak n} \lambda/2$,
and vanishes at nodes, $x = (1/2 \pm {\mathfrak n})\lambda/2$,
where ${\mathfrak n}=0,1,2,\ldots$
The electron phase space is 6-dimensional, $(x,y,z,p_x,p_y,p_z)$,
however the coordinates $y$ and $z$ are {\it ignorable} (do not influence the dynamics),
because the EM field depends only on $(t,x)$.
In a LP SW where the electric field is polarized along $z$-axis,
the equation for the  $y$-component of the electron momentum
contains only dissipative terms, thus $p_y$ is exponentially dumped, in general.
This allows us to restrict the presentation of the electron trajectories
to the phase subspace $(x,p_x,p_z)$ for LP SW,
and to $(x,p_y,p_z)$ for CP SW.

\begin{figure}[H]
\includegraphics{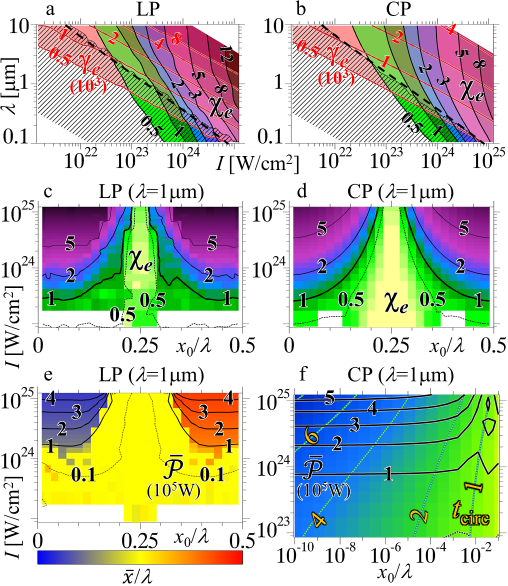}
\caption{%
(color).
The maximum Lorentz factor, $\gamma_e$ (red curves)
and the maximum $\chi_e$ parameter (black curves) and 
escaping trajectories (hatched region)
in the linearly (a, LP) and
circularly (b, CP) polarized standing waves (SW) 
versus the wave intensity, I, and wavelength, $\lambda$.
Dashed lines for $I\times(\lambda/1\mu{\rm m})^{1.6} = 2\times 10^{23}$ W/cm$^2$.
The maximum $\chi_e$ parameter (curves and colorscale)
on the trajectories originated at $x_0$ 
in the LP (c) and CP (d) SW for $\lambda=1\mu{\rm m}$.
The time-averaged longitudinal coordinate ($\bar{x}$, colorscale)
of the trajectories originated at $x_0$
in the LP SW (e) and the duration of a circular motion
($t_{\rm circ}$, in wave periods, dashed curves and colorscale)
for CP SW (f), for $\lambda=1\mu{\rm m}$.
Black curves in (e), (f) for the time-average emitted power in $10^5$ W.
}
\label{fig-2:max-chi-gamma}
\end{figure}

Fig. \ref{fig-2:max-chi-gamma} shows data for electrons moving in a LP or CP EM SW
at different intensity, $I$, and wavelength, $\lambda$,
varying as follows:
$I$ from $10^3 I_0 = 1.37\times 10^{21}$ 
to $10^7 I_0 = 1.37\times 10^{25}$W/cm$^2$
and $\lambda$ from $0.1$ to $10\,\mu$m.
Electrons with zero momentum are initially located 
within the SW spatial period, $0<x_0/\lambda<1/2$.
They remain confined in this interval or escape,
depending on the EM SW intensity and wavelength.
Hatched regions in  Fig. \ref{fig-2:max-chi-gamma}(a,b)
indicate $I$ and $\lambda$, at which escaping electrons appear.
Such electrons perform a ``random walk'' motion:
they travel over large distances
migrating between SW spatial periods
sometimes oscillating near the electric field node planes
\cite{Gonoskov,RelAstr-PPR}.
The boundary between the regions of escaping and confined trajectories
roughly corresponds to the curve $a=0.5 a_{\rm RQ}$ in Fig. \ref{fig-1:I-lambda-zones}.

When the RR becomes significant,
the  dissipation contracts the possible volume occupied by trajectories,
resulting in trapping of electrons within the SW spatial period.
Fig.  \ref{fig-2:max-chi-gamma}(a,b)
presents the maximum Lorentz factor, $\gamma_e$,
and the maximum $\chi_e$ parameter
reached by electrons on their trajectories.

\begin{figure}[H]
\includegraphics{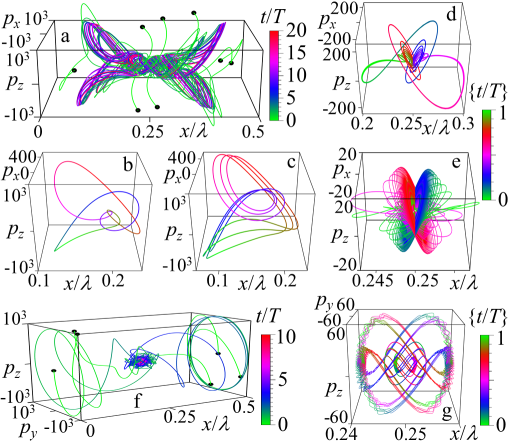}
\caption{%
(color).
Typical electron trajectories and attractors 
in linear (a-e) and circular (f,g) polarized standing wave 
for $I=1.37\times 10^{24}$ W/cm$^2$, $\lambda=1\mu{\rm m}$.
Black dots (a,f) for initial locations. 
Color for time in wave periods;
in (b-e,g) {} denote the fractional part.
}
\label{fig-3:attractors}
\end{figure}

Fig. \ref{fig-2:max-chi-gamma}(c-f) shows the characteristics of confined trajectories
(white regions indicate escaping trajectories).
The maximum $\chi_e$ parameter
obtained on the trajectories originated at $x_0$
is shown in Fig.  \ref{fig-2:max-chi-gamma}(c,d)
for the SW wavelength of $\lambda=1\;\mu$m.
Trajectories starting near the electric field node exhibit lower $\chi_e$ values.
Frame (e) shows the time-averaged longitudinal coordinate, $\bar{x}$,
of the trajectories originated at $x_0$ in the LP SW.
Electrons fall into few attractors near the electric field nodes and antinodes,
where they perform quasi-periodic oscillations.
The same style frame for the CP SW would be homogeneous in color
since all the trajectories in the CP SW, if they are confined in the SW spatial period,
reduce to low-magnitude oscillations near the electric field node (see Fig. \ref{fig-3:attractors}).
Therefore in (f) we present the duration of a circular motion, $t_{\rm circ}$,
for trajectories originated very close to the electric field antinode.
The maximum $\gamma_e$ and $\chi_e$ distribution on the $(I,\lambda)$ plane
for an asymptotic motion of confined electrons is almost the same as (a) for LP SW
and very much different form (b) for CP SW, where the asymptotic motion is weakly relativistic.
The time-average power of EM radiation 
emitted by an electron on its trajectory
is shown in (e,f) by black curves. 
For LP SW (e) it is for an asymptotic motion
while for CP SW (f) it is for the initial circular motion
near the electric field antinode.
The higher the CP SW amplitude, the longer the circular motion.


Typical electron trajectories in the phase space for the LP and CP SW
are exemplified in Fig. \ref{fig-3:attractors},
for $I=1.37\times 10^{24}$ W/cm$^2$, $\lambda=1\;\mu$m.
Due to a strong dissipation, electrons started with different momenta,
Fig. \ref{fig-3:attractors}(a,f),
quickly fall to an asymptotic motion near attractors.
In the LP SW, Fig. \ref{fig-3:attractors}(a),
electrons fall into different attractors depending on their 
initial coordinate and momentum (as was seen in Fig. \ref{fig-2:max-chi-gamma}(e)).
In the CP SW, Fig. \ref{fig-3:attractors}(f),
electrons fall into a single attractor at the electric field node
irrespective of the initial coordinate and momentum.
However, if electrons where initially sufficiently close to the  electric field antinode,
then before falling into the attractor 
they perform few rotations emitting all the power received from the CP SW,
as seen in Fig. \ref{fig-2:max-chi-gamma}(f).

In the LP SW, near the electric field antinode 
there are limit cycles Fig. \ref{fig-3:attractors}(b,c),
characterized by a large values of $\gamma_e$, $\chi_e$
and time-averaged emitted power, Fig. \ref{fig-2:max-chi-gamma}(a,c,e).
Near the electric field node
we see a bow-like limit cycle Fig. \ref{fig-3:attractors}(d)
with relatively high $\chi_{e\,\rm max}\sim 3$,
and a knot-like attractor Fig. \ref{fig-3:attractors}(e),
much weaker in terms of $p_x, p_z$.
Due to the symmetry of motion equations,
a confined trajectory has a twin which is antisymmetric in the phase space
with respect to the electric field node.
The oscillations near the electric field node
exhibit substantial frequency upshift,
so that electron trajectory makes many cycles
around the fixed point $x/\lambda=1/4$
during one time cycle of the LP SW.
The analysis of linearized motion equations
shows that in this case
the electron dynamics is characterized by two timescales,
one corresponding to the SW frequency
and another being determined by the SW amplitude
and the electron Lorentz factor.

In the CP SW, Fig. \ref{fig-3:attractors}(f),
all the trajectories fall into one of 
an infinite system of limit cycles
near the electric field node,
which forms a pompom-like structure Fig. \ref{fig-3:attractors}(g).
Here $\chi_e \ll 1$.

The kind and localization of attractors
strongly depend on the SW intensity and wavelength.
For the same intensity of $I=1.37\times 10^{24}$ W/cm$^2$,
the case of $\lambda=3\,\mu$m shows different limit cycles
near the electric field antinodes
but the same type attractors near nodes.
When $\lambda=0.3\,\mu$m,
the classical RR force is strongly weakened by the QED effects,
so the confined trajectories chaotically 
jump between different types of oscillations
resembling motion near attractors seen at higher $\lambda$.


The attractors near the electric field node in LP and CP SW
exhibit properties of strange attractors,
playing a fundamental role in the theory of dynamic systems \cite{EckmannRuelle}.
As numerical analysis shows,
the trajectories near the electric field node behave as dense periodic orbits.
The trajectories in the LP SW 
starting with a zero momentum
from $x_0$ corresponding to the time-average $\bar{x} \approx 1/4$,
yellow region in Fig. \ref{fig-2:max-chi-gamma}(e),
and trajectories in the CP SW
starting near $x_0=1/4$
are highly sensitive to the initial coordinate.
More precisely, the indication of the presence of strange attractors
is given by the maximal Lyapunov exponent \cite{EckmannRuelle},
\begin{equation}
\Lambda=\lim_{t \to \infty} t^{-1} \lim_{ \delta_0 \to 0}
               \ln[\delta(t) / \delta_0],
\label{Lyapunov}
\end{equation}
where $\delta(t)$ is the distance in the phase space
between the ends of two trajectories,
initially separated by $\delta_0 = \delta(0)$.
For the attractor in the LP SW, shown in Fig. \ref{fig-3:attractors}(e),
the maximum Lyapunov exponent numerically estimated
as $\Lambda\gtrapprox 5$,
and for the CP SW case, shown in Fig. \ref{fig-3:attractors}(g),
as $\Lambda\gtrapprox 1$.
Since they are both positive,
the trajectories confined near the electric field node
are chaotic, revealing strange attractors.


According to our numerical analysis,
confined motion of electrons 
near the electric field nodes and antinodes
is robust with respect to small imbalances of the 
intensity and wavelength of EM waves forming the SW.
Even though the nodes and antinodes are no longer stationary in this case,
the asymptotic motion resembles oscillations near attractors,
slowly drifting along the $x$-axis during an oscillation cycle.

\begin{figure}[H]
\includegraphics{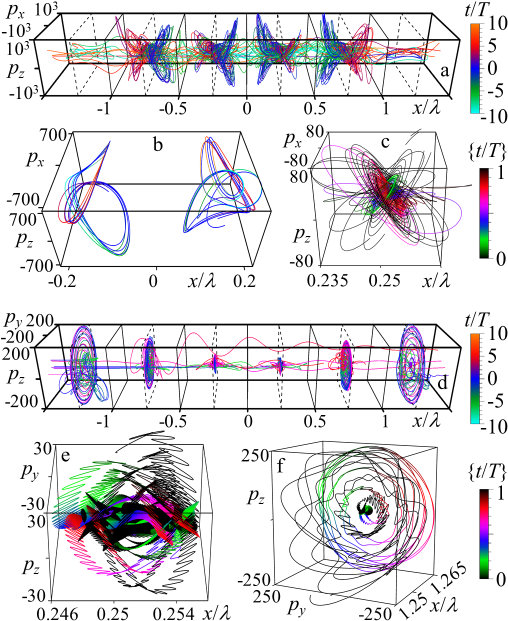}
\caption{%
(color).
The electron trajectories in the standing wave 
formed by two colliding 
linearly (a,b,c, LP) and circularly (d,e,f, CP) polarized
laser pulses
propagating along $x$-axis and having 
$I=1.37\times 10^{24}$ W/cm$^2$, $\lambda=1\mu{\rm m}$,
duration 33 fs and focal spot 3 $\mu{\rm m}$.
}
\label{fig-4:paraxial}
\end{figure}

This robustness also manifests itself when a transient standing wave
is formed by two counter-propagating paraxial gaussian laser beams
with the intensity $I=(1.37/4)\times 10^{24}$ W/cm$^2$,
wavelength $\lambda=1\;\mu$m,
duration $10\lambda/c = 33$ fs and focal spot 3 $\mu$m,
Fig. \ref{fig-4:paraxial}.
The constructive interference of  these two paraxial pulses
gives the SW peak intensity of $I=1.37\times 10^{24}$ W/cm$^2$.

Initially (at $t=-20\lambda/c$) randomly  distributed in the box $[-1.5,1.5]^3 \lambda^3$,
$10^3$ electrons move in the transient SW,
which is formed around $t=0$ for approximately $3\lambda/c$.
About 10\% and 15\% electrons remain in the box 
for more than 25 laser cycles in the LP and CP SW, respectively.
Their trajectories are exemplified in Fig. \ref{fig-4:paraxial}(a,d).
These trajectories fall into attractors typical for a 1D SW,
although somewhat displaced due to non-stationary location of peripheral nodes and antinodes.
In the LP case, we see limit cycles near the electric field nodes
and a knot-like attractors at antinodes, Fig. \ref{fig-4:paraxial}(b,c).
In the CP case, we see pompom-like attractors at antinodes, Fig. \ref{fig-4:paraxial}(e,f).


In conclusion,
in the electron dynamics in a strong electromagnetic standing wave,
the electromagnetic wave intensity $I_{\rm RQ} \sim 10^{24}$ W/cm$^2$
and wavelength $\lambda_{\rm RQ} \sim 1\,\mu$m,
reachable in near future by high-power lasers
\cite{ELI-Beamlines},
separate different regimes of radiation reaction.
For $I<I_{\rm RQ}$ and $\lambda>\lambda_{\rm RQ}$,
it is possible to achieve strong classical RR force without significant QED effects,
whereas 
for $I>I_{\rm RQ}$ and $\lambda<\lambda_{\rm RQ}$,
strong QED effects can be seen with a relatively small classical RR force.

When radiation reaction is significant,
a strong dissipation results in formation of limit cycles and strange attractors,
depending on the standing wave polarization.
A high energy and $\chi_e$ parameter
are achieved on limit cycles near the electric field antinode
in the linearly polarized standing wave.

When a transient standing wave is formed
by two intense counter-propagating laser pulses,
a substantial number of electrons is confined
for the standing wave lifetime
($\approx1/3$ of a Gaussian laser pulse duration),
tracing limit cycles and strange attractors
peculiar to ideal planar standing wave.
Since electrons forget their initial momenta due to dissipation,
the interaction of a GeV electron beam transversely
propagating through a collision point of two 
multi-petawatt counter-propagating laser pulses
can create a {\it transient microscopic storage ring}
similar to a synchrotron,
which reveals itself by a characteristic transverse
high-power high-frequency electromagnetic radiation
and  the resulting spatial and spectral electron distribution.
This creates a new framework
for high energy physics experiments,
in particular, 
solving a well-known problem of 
charged particles delivery to the highest intensity
region of the electromagnetic field.

\begin{acknowledgments}
We thank 
M. Jirka,
O. Klimo,
S. Weber,
and A. G. Zhidkov
for discussions. 
S.V.B. and T.Z.E. acknowledge support from ELI-Beamlines.
\end{acknowledgments}

\end{document}